\documentclass[aps,12pt]{revtex4}
\usepackage{amsmath}
\usepackage{amsfonts}
\usepackage{amssymb}
\usepackage{color}
\usepackage{graphicx}
\usepackage{mathrsfs}

\def\bc{\begin{center}}
\def\nno{\nonumber}
\def\ec{\end{center}}
\def\be{\begin{eqnarray}}
\def\ee{\end{eqnarray}}

\definecolor{dyellow}{rgb}{1.,0.8,.0}
\definecolor{myblue}{rgb}{.1,.1,.7}
\definecolor{dcyan}{rgb}{.0,.6,.6}
\definecolor{dmagenta}{rgb}{0.6,0.0,0.6}
\definecolor{brown}{rgb}{0.6,0.2,0.}
\definecolor{darkblue}{rgb}{.0,.0,0.5}
\definecolor{darkred}{rgb}{0.75,0.0,0.0}
\definecolor{orange}{rgb}{1.,.6,.0}
\definecolor{dorange}{rgb}{0.8,.4,.0}
\definecolor{darkgreen}{rgb}{0.0,0.6,0.0}
\definecolor{purple}{rgb}{.4,.0,.4}
\definecolor{lightgrey}{rgb}{0.7, 0.7, 0.7}
\definecolor{grey}{rgb}{0.4, 0.4, 0.4}

\def\dl{\delta}
\def\eps{\epsilon}
\def\ka{\kappa}
\def\la{\lambda}

\def\si{\sigma}
\def\om{\omega}

\def\pa{\partial}

\def\Dl{\Delta}
\def\La{\Lambda}

\def\Om{\Omega}

\newcommand\rha{\rightarrow}

\def\PRD{{Phys. Rev.}~{\bf D}}

\def\hl{Ho\v{r}ava-Lifshitz~}

\begin{document}

\title{Holographic Superconductors with Ho\v{r}ava-Lifshitz Black Holes}
\vskip 2cm \vskip 2cm
\author{Rong-Gen Cai}\email{cairg@itp.ac.cn}
\author{Hai-Qing Zhang}\email{hqzhang@itp.ac.cn}
\address{Key Laboratory of Frontiers in Theoretical Physics,
Institute of Theoretical Physics, Chinese Academy of Sciences,
   P.O. Box 2735, Beijing 100190, China}


\begin{abstract}
 We discuss the phase transition of planar black holes in Ho\v{r}ava-Lifshitz gravity
 by introducing a Maxwell field and a complex scalar field. We calculate
the condensates of the charged operators in the dual CFTs when the
mass square of the complex scalar filed is $m^2=-2/L^2$ and
$m^2=0$, respectively. We compute the electrical conductivity of
the \hl superconductor in the probe approximation. In particular,
it is found that there exists a spike in the conductivity for the
case of the operator with scaling dimension one. These results are
quite similar to those in the case of Schwarzschild-AdS black
holes, which demonstrates that the holographic superconductivity
is a robust phenomenon associated with asymptotic AdS black holes.

\end{abstract}


\maketitle

\section{Introduction}
  The AdS/CFT correspondence~\cite{maldacena,polyakov,witten} relates a weak
 coupling gravity theory in an anti-de Sitter space
 to a strong coupling conformal field theory in one less dimensions.
 Recently it has got  a lot of applications in condensed matter
 physics~\cite{hart,herzog}, in particular in understanding some phenomena such as
  superconductivity~\cite{gubser,hhh} and
 superfluid~\cite{basu,herzog2}. It turns out that a simply
 Einstein-Maxwell theory coupled to a charged complex scalar field
 could provide a holographic description of a superconductor.

  The hairy black hole with a complex scalar
 field~\cite{gubser08} can be explained as a charged condensation in the dual CFTs
 according to the AdS/CFT correspondence. The expectation value of charged operators
 in the dual CFTs will break U(1) symmetry; According to
 Weinberg~\cite{wein}, this produces a superconducting phase. In
 the paper~\cite{hhh}, the authors considered the
 holographic superconductors in the Abelian-Higgs model within the probe limit,
  which means the backreaction of the dynamical matter field on the
  spacetime metric is negligible. The probe limit is justified
  when the charge of the complex scalar field is large. For the
  Einstein-Abelian-Higgs system, one has a consistent solution: a
  Schwarzschild-AdS black hole solution with a constant scalar
  field. When the black hole temperature $T$ decreases and is below a critical
 temperature $T_c$, it is found that the Schwarzschild-AdS black hole becomes
  unstable,
 and a new black hole solution with charged scalar hair is favored.
 It implies that a second order phase transition happens from the Schwarzschild-AdS black hole
 to the hairy black hole.  In the field theory side, it indicates
 the condensation of the charged operator comes out and a
 superconducting phase occurs. According to the AdS/CFT correspondence,
 the Maxwell field in the bulk is dual to a $U(1)$ conserved current in the
 boundary field theory. Therefore, considering the fluctuation of the Maxwell field,
 one is able to calculate the conductivity. The authors of \cite{hhh} found
 that the behaviors of the condensation and the conductivity are
 qualitatively consistent with the BCS theory~\cite{parks}.

  The backreaction effect of matter field has been studied in~\cite{hhh2},
   the phase diagram and conductivity are also still
  consistent with the BCS theory, as expected. Extensions of the holographic
  superconductor to higher spin case have been done in~\cite{gubser0803,rh,gp}.
  The interesting holographic superconductor models have been
  embedded in string/M theory~\cite{dh,gubser3,gsw,gpr}. More recently,
  the investigation of the ground state (zero-temperature limit) of
  holographic superconductors has revealed that
  there may exist some Lifshitz scaling symmetry for the bulk
  solutions and bulk solutions have vanishing entropy~\cite{gsw,gubser4,hr2,kz}.

 Recently some interest arises to build the holographic superconductors
 in the bulk backgrounds with Lifshitz/Schr\"{o}dinger scaling
 symmetry, according to the AdS/NCFT (non-relativistic conformal field)
 correspondence~\cite{bryn,zhou,cmo}.  On the other hand, more recently a lot of attention has been
 focused on \hl theory~\cite{ho}, which is a power accounting
 renormalizable gravity theory.  In particular, the \hl  theory is a
 non-relativistic theory. Note that the fact that
 superconductivity is a phenomenon which can be described by a
 non-relativistic field theory. It is therefore of interest to
 study the holographic superconductivity in the \hl theory,
 partially in order to see how the phenomenon of the charged
 condensation is robust, and partially to see what difference will
 appear, compared to the case of the relativistic general relativity.

 This paper is organized as follows. In Sec.~\ref{sect:rud} we briefly
 introduce the \hl gravity including its action, planar black hole solution
   and associated thermodynamics. Sec.~\ref{sect:conden} considers the phase
 transition of the \hl holographic superconductors by introducing
a complex scalar field and Maxwell field into the planar black
hole background.  The condensate of the charged operators
 is calculated numerically. In Sec.~\ref{sect:conduc}, we
 calculate the electrical conductivity in the probe approximation. We find
 that the behavior of the conductivity is qualitatively
 consistent with the BCS theory and a spike will appear in the
 conductivity if the operator is of scaling dimension one.
 Employing the method in \cite{hr2}, we try
 to qualitatively explain this phenomenon using the one-dimensional
 Schr\"{o}dinger equation. We conclude our paper in Sec.~\ref{sect:con}.

\section{Black holes in  Ho\v{r}ava-Lifshitz gravity}
\label{sect:rud} In non-relativistic field theory, space and time
have different scalings, which is called anisotropic scaling,
 \be x^i\rightarrow bx^i, \qquad t\rightarrow b^zt,\qquad i=1,2,3.\ee
where $z$ is called {\it dynamical critical exponent}.  In order
to present the non-relativistic \hl gravity, it turns out
convenient to employ the ADM formalism
 \be
ds^2=-N^2dt^2+g_{ij}(dx^i+N^idt)(dx^j+N^jdt).
 \ee
 where $N$ is the lapse function, $N^i$ is the shift vector, and $g_{ij}$ is the metric of
the spacelike hypersurface. The extrinsic curvature for the
hypersurface is
  \be
 K_{ij}=\frac{1}{2N}(\dot{g_{ij}}-\nabla_iN_j-\nabla_jN_i),
  \ee
where the dot denotes the derivative with respect to time $t$, and
$\nabla_i$ stands for the covariant derivative associated with the
metric $g_{ij}$. In order for a theory to be power counting
renormalizable, the critical exponent has at least $z=3$ in four
spacetime dimensions. The action of the \hl theory with $z=3$
is~\cite{ho}
 \be
  S=&& \int dtd^3x \sqrt{g}N
  \big[\frac{2}{\ka^2}(K_{ij}K^{ij}-{\tilde{\la}}
 K^2)+\frac{\ka^2\mu^2(\La
 R-3\La^2)}{8(1-3\tilde{\la})}+\frac{\ka^2\mu^2(1-4\tilde{\la})R^2}{32(1-3\tilde{\la})}\\\nno
 &&-\frac{\ka^2\mu^2}{8}R_{ij}R^{ij}+\frac{\ka^2\mu\eps^{ijk}}{2w^2}R_{il}\nabla_jR^l_{\
 k}-\frac{\ka^2}{2w^4}C_{ij}C^{ij}\big]
 \ee
where $\ka^2,\tilde{\la}, \mu, w^2$ are the coupling constants and
$\La$ is related to the effective cosmological constant; $C_{ij}$
is the Cotten tensor, defined as
 \be C^{ij}=\eps^{ikl}\nabla_k(R^j_{\ l}-\frac1
 4R\dl^j_l). \ee
In the \hl theory, the topological black hole solutions with
constant scalar curvature horizon in the case of $\tilde{\la}=1$ are
presented in~\cite{cai}; the metric is of the form
  \be\label{line} ds^2=-N_0^2f(r)dt^2+\frac{dr^2}{f(r)}+r^2d\Om^2_k.\ee
 where $N_0$ is a constant, which can be set to the unity, $d\Om^2_k$ is a 2-dimensional Einstein space
 with constant scalar curvature $2k$ (without loss of generality, $k$ can be taken  $0$ and
 $\pm1$). The metric function $f$ is given by
  \be f(r)=k+\tilde x^2-\sqrt{c_0\tilde x}.
  \ee
 where $\tilde x=\sqrt{-\La}~r$, $c_0$ is an integration constant and
 $c_0>0$. The mass of this topological black hole solution is
  \be M=\frac{\ka^2\mu^2\sqrt{-\La}\Om_k}{16}c_0, \ee
 where $\Om_k$ is the volume of the 2-dimensional Einstein space.
 And the Hawking temperature of the black hole is
  \be T=\frac{3\tilde x_+^2-k}{8\pi \tilde x_+}\sqrt{-\La}. \ee
 where $\tilde x_+$ is the largest root of $f(r)=0$.

\section{The condensate of charged operators}
\label{sect:conden} We start with the planar \hl black holes with
$k=0$. In that case the metric
 (\ref{line}) can be written as
 \be\label{metric}  ds^2=-f(r)dt^2+\frac{dr^2}{f(r)}+r^2(dx^2+dy^2),\ee
with
 \be\label{f} f=\tilde x^2-\sqrt{c_0\tilde x}. \ee
One can see that in the large $r$ limit or when $ c_0=0$, the
spacetime becomes a four dimensional anti-de Sitter space,
AdS$_4$. Note that the solution (\ref{metric}) has very unusual
asymptotic behavior, compared to the Schwarzschild-AdS black hole.
Comparing with the standard AdS$_4$ spacetime, we may set
$-\La=\frac{1}{L^2}$, where $L$ is the radius of AdS$_4$.

The Lagrangian for the system of the Maxwell field and the complex
scalar field is of the Abelian-Higgs type~\cite{hhh}
 \be\label{lag} \mathcal{L}=-\frac1 4
 F^{\mu\nu}F_{\mu\nu}-|\nabla_{\mu}\psi-iqA_{\mu}\psi|^2-V(|\psi|),
 \ee
where $F^{\mu\nu}$ is the Maxwell field strength $F=dA$ and $\psi$
is the complex scalar field with the potential $V$. We have from
(\ref{lag}) the equation of the scalar field
 \be \label{seq}
 -(\nabla_{\mu}-iqA_{\mu})(\nabla^{\mu}-iqA^{\mu})\psi+ \frac{\pa V(|\psi|)}{\pa\psi^*}=0 ,\ee
while the Maxwell's equation reads
 \be\label{meq}
 \nabla^{\mu}F_{\mu\nu}=iq\big(\psi^*(\nabla_{\nu}-iqA_{\nu})\psi-\psi(\nabla_{\nu}+iqA_{\nu})\psi^*\big).\ee
Consider the ansatz
 \be A_{\mu}=(\phi(r),0,0,0),\quad \psi=\psi(r).\ee
This ansatz implies that the phase factor of the complex scalar
field is a constant. Thus one may take $\psi$ to be real. In the
black hole background (\ref{metric}), the equation of scalar field
(\ref{seq}) then reduces to
 \be\label{seq2} \psi''+(\frac{f'}{f}+\frac2
 r)\psi'+\frac{q^2\phi^2}{f^2}\psi-\frac{1}{2f}\frac{\pa V(\psi)}{\pa\psi}=0,\ee
 where  a prime denotes the derivative  with respect to $r$, and the Maxwell's equations is simplified to
 \be\label{meq2} \phi''+\frac2 r
 \phi'-\frac{2q^2\psi^2}{f}\phi=0.\ee
At the black hole horizon $r=r_+$, one must have $\phi=0$ because
its norm is required finite, and the scalar field  $\psi$ should
also be finite there, namely, $\psi|_{r=r_+}<+\infty$.

In order to get the explicit behavior of $\phi(r)$ and $\psi(r)$
on the boundary $r\rha \infty$. We  assume the potential to be
$V(\psi)=m^2\psi^2$ and consider two cases, one is  $m^2L^2=-2$
and the other is $m^2=0$; both are above the
Breitenlohner-Freedman bound~\cite{bf}. From now on, we set the
radius of AdS$_4$ to be $L=1$.  Let us first discuss the case with
$m^2=-2$. From (\ref{seq2}) and (\ref{meq2}), we can easily get
their behavior in the large $r$ limit,
 \be \label{sasy}
 \psi|_{r\rha\infty}=\frac{\psi^{(1)}}{r}+\frac{\psi^{(2)}}{r^2}+\cdots,\ee
and
 \be\label{masy} \phi|_{r\rha\infty}=\mu-\frac{\rho}{r}+\cdots,\ee
 where  $\mu$ is the
chemical potential and $\rho$ is the charge density on the
boundary. Because the boundary is a (2+1)-dimensional field
theory, $\mu$ is of mass dimension one and $\rho$ is of mass
dimension two.  From the boundary behaviors, we can read off the
expectation values of operator $\mathcal{O}$ dual to the field
$\psi$. From \cite{kw}, we know that for $\psi$, both of these
falloffs are normalizable, and in order to keep the theory stable,
we should either impose
 \be \psi^{(1)}=0,\quad \text{and} \quad \langle
 \mathcal{O}_2\rangle=\sqrt{2}\psi^{(2)},\ee
or
 \be \psi^{(2)}=0,\quad \text{and} \quad \langle
 \mathcal{O}_1\rangle=\sqrt{2}\psi^{(1)}.\ee
The factor of $\sqrt{2}$ is a convenient normalization as in
\cite{hhh}. The index $i$ in $\psi^{(i)}$ represents the scaling
dimension $\la_{\mathcal{O}}$ of its dual operator
$\langle\mathcal{O}_i\rangle$, {\it i.e.}, $\la_{\mathcal{O}_i}=i
$.

 We now find the numerical solutions of Eqs.~ (\ref{seq2}) and
 (\ref{meq2}) with the boundary conditions mentioned above. Because
 the dimension of temperature $T$ is of mass dimension one, the
 ratio $T^2/\rho$ is dimensionless. Therefore
 increasing $\rho$, while $T$  is fixed,  is equivalent to decrease $T$ while $\rho$ is
 fixed. In our calculation, we find that when $\rho>\rho_c$, the operator
 condensate will appear; this means when $T<T_c$ there will be an
 operator condensate, that is to say, the superconducting phase
 occurs.
 \begin{figure}[htb]\label{conden}
\includegraphics[scale=0.76]{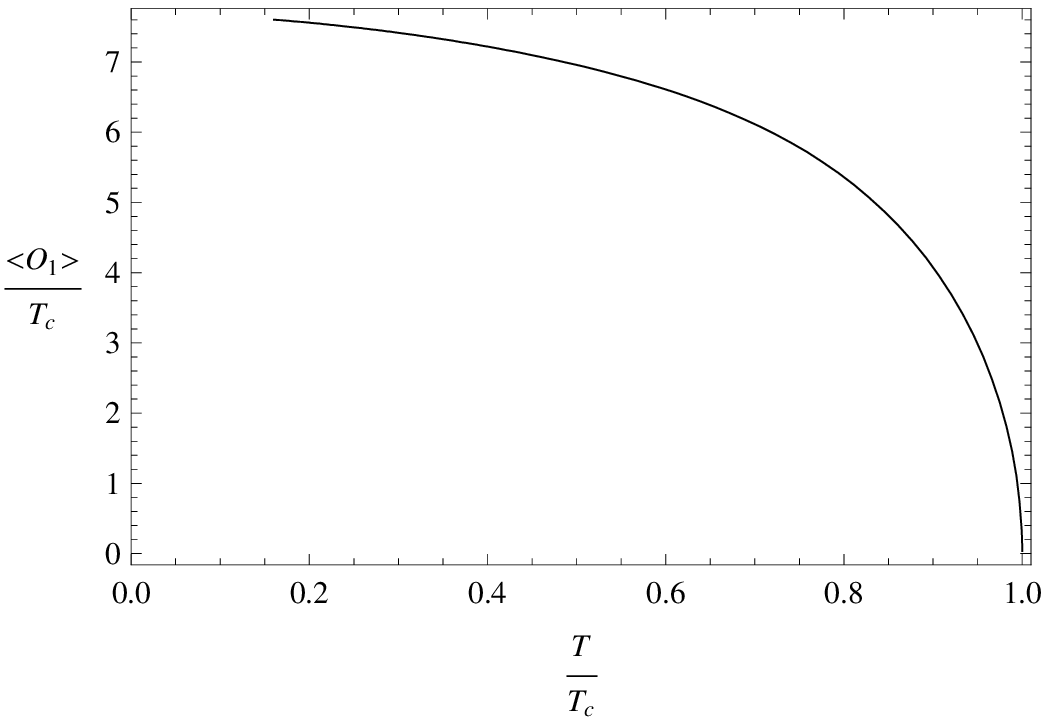}
\includegraphics[scale=0.76]{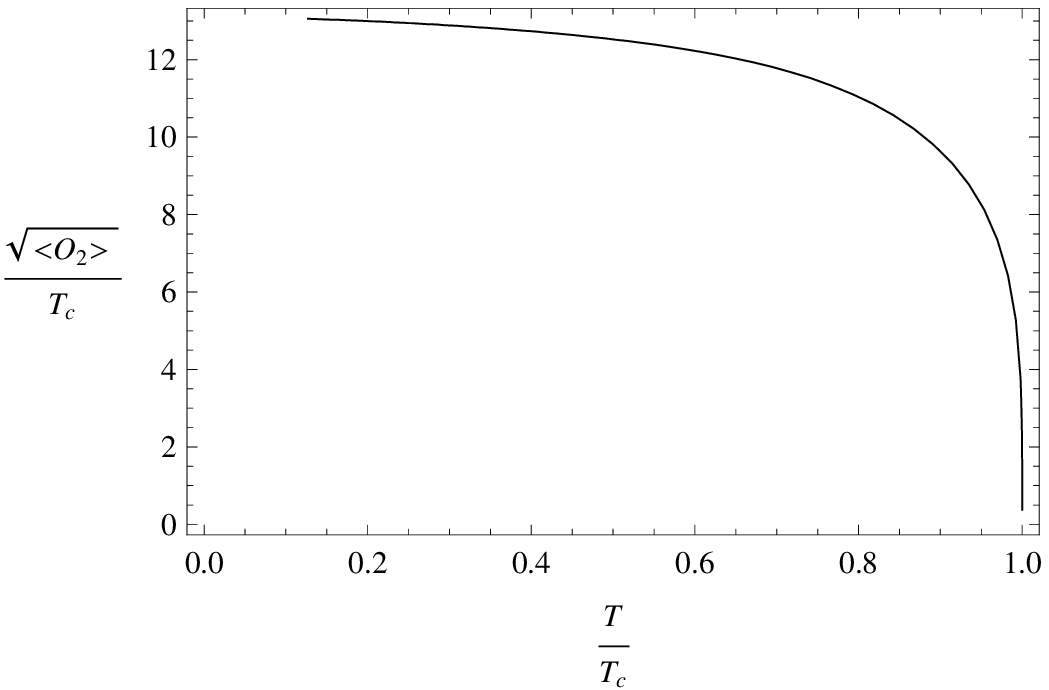}
\caption{The condensates of operators $\mathcal{O}_1$(left) and
$\mathcal{O}_2$(right) versus temperature. The condensates
disappear as $T\rha T_c$.}
\end{figure}

We plot in Fig.~1 the condensates of operators $\mathcal{O}_1$ and
$\mathcal{O}_2$.  They approach to fixed constants as $T\rha 0$,
which is qualitatively consistent with that obtained in BCS
theory. In the  BCS theory, the expectation values of $\langle
\mathcal{O}_1 \rangle$ and $\sqrt{\langle \mathcal{O}_2\rangle}$
are twice the superconducting gap.  The theoretical prediction for
the expectation value is $2\times \text{gap}=3.54T_c$ at
$T=0$~\cite{tinkham}. In our holographic model, however, we see
from Fig.~1 that the expectation values of the two operators are
higher than $3.54T_c$. This may be explained that the holographic
model describes a strong interacting theory than the BCS
theory~\cite{herzog}.

 In the mean field theory, near $T\sim T_c$, the order parameter
 has the behavior
   \be \langle
   \mathcal{O}_1\rangle\sim\langle\mathcal{O}_2\rangle\sim(T_c-T)^{1/2}.\ee
In our case,  we can fit the data near $T\rha T_c$ and find
 \be \langle \mathcal{O}_1\rangle \approx 13.17 T_c
 \sqrt{1-T/T_c},\quad \text{when} \quad T\rha T_c, \ee
where $T_c\approx 0.0406243 \sqrt{\rho}$, while
 \be \langle \mathcal{O}_2\rangle \approx 304.542T_c^2\sqrt{1-T/T_c}, \quad  \text{when}\quad T\rha T_c.\ee
where $T_c\approx 0.0762589\sqrt{\rho}$. These results are
qualitatively the same as the case for Schwarzschild-AdS black
hole~\cite{hhh}.

To see the condensate dependence on the mass of the scalar field,
 we will consider the case with
$m^2=0$.
 In that case, the asymptotic
behavior of scalar field near the boundary is
 \be \psi|_{r\rha
 \infty}=\psi^{(0)}+\frac{\psi^{(3)}}{r^3}+\cdots,
 \ee
 where
$\psi^{(0)}$ is non-normalizable, which therefore we will not
consider any more and $\psi^{(3)}$ is now proportional to the
expectation value of the dual operator $\mathcal{O}_3$, whose
scaling dimension is $3$.
Near the critical temperature, we find
 \be \langle \mathcal{O}_3\rangle=5166.193
 T_c^3\sqrt{(1-T/T_c)},\quad \text{when}\quad T\rha T_c.\ee
where $T_c\approx 0.04988 \sqrt{\rho}$.  To compare different
behaviors of condensate for operators with different scaling
dimensions, we plot the condensate of the operator $\mathcal{O}_3$
with the other two in Fig~2. It shows that the condensate
increases when the scaling dimension $\la$ gets larger, which is
consistent with the argument given in~\cite{hr}.
\begin{figure}[htb]\label{condenall}
\includegraphics[scale=0.9]{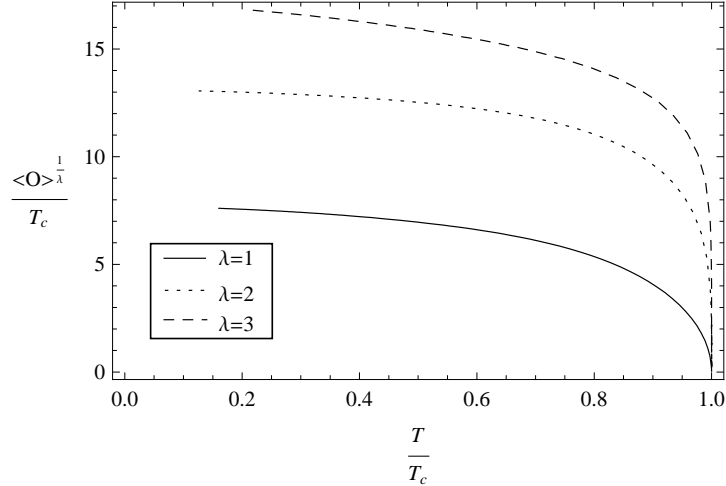}
\caption{The condensate of three operators with scaling
dimensions, which are labelled by $\la$.  It can be seen that the
condensate increases when $\la$ gets larger. }
\end{figure}

\section{The electrical conductivity}
\label{sect:conduc}
  In order to compute the electrical conductivity, according to
  the AdS/CFT dictionary, we perturb the Maxwell field and the time-space
component of the black hole background,  then calculate the linear
response to the perturbation.  In the probe approximation, the
effect of the perturbation of metric can be ignored. In addition,
we assume that the perturbation of the vector potential is
 translational symmetric and has a time dependence as  $\dl A_x=A_x(r)e^{-i\om
 t}$. Substitute this perturbation into Maxwell's equations (\ref{meq}), we
 have
  \be \label{pert}
  A_x''+\frac{f'}{f}A_x'+(\frac{\om^2}{f^2}-\frac{2q^2\psi^2}{f})A_x=0.\ee
 At the black hole horizon, the ingoing wave condition should be employed, $A_x\propto
 f^{-i\om/3r_+}$, while at the infinity boundary, the asymptotic behavior of $A_x$
 is
 \be \label{asym} A_x=A_x^{(0)}+\frac{A_x^{(1)}}{r}+\cdots \ee
In AdS/CFT correspondence, it is well-known that $A_x^{(0)}$
corresponds to  the source and $A_x^{(1)}$ to the expectation
value for the current in the dual CFT,
 \be A_x=A_x^{(0)},\quad \langle J_x\rangle=A_x^{(1)}.\ee
By the Ohm's law, the conductivity can be calculated as
 \be \label{ohm} \si(\om)=\frac{\langle J_x\rangle}{E_x}=-\frac{i\langle J_x\rangle}{\om
 A_x}=-\frac{i A_x^{(1)}}{\om A_x^{(0)}}.\ee

\begin{figure}[htb]\label{conduc}
\includegraphics[scale=0.76]{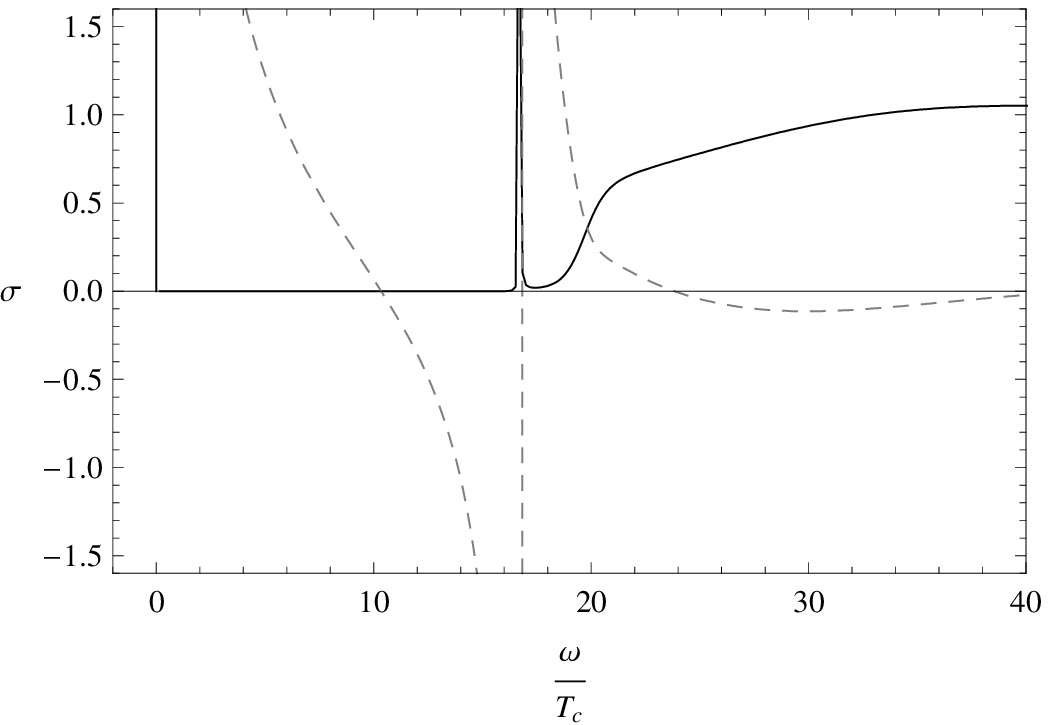}
\includegraphics[scale=0.76]{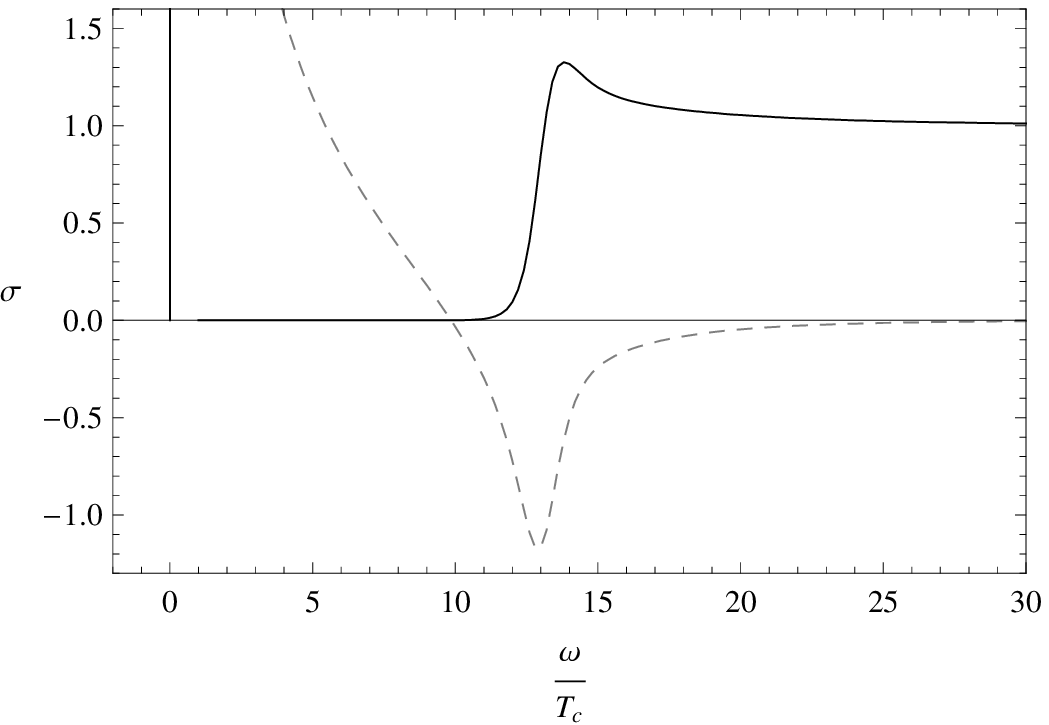}
\includegraphics[scale=0.76]{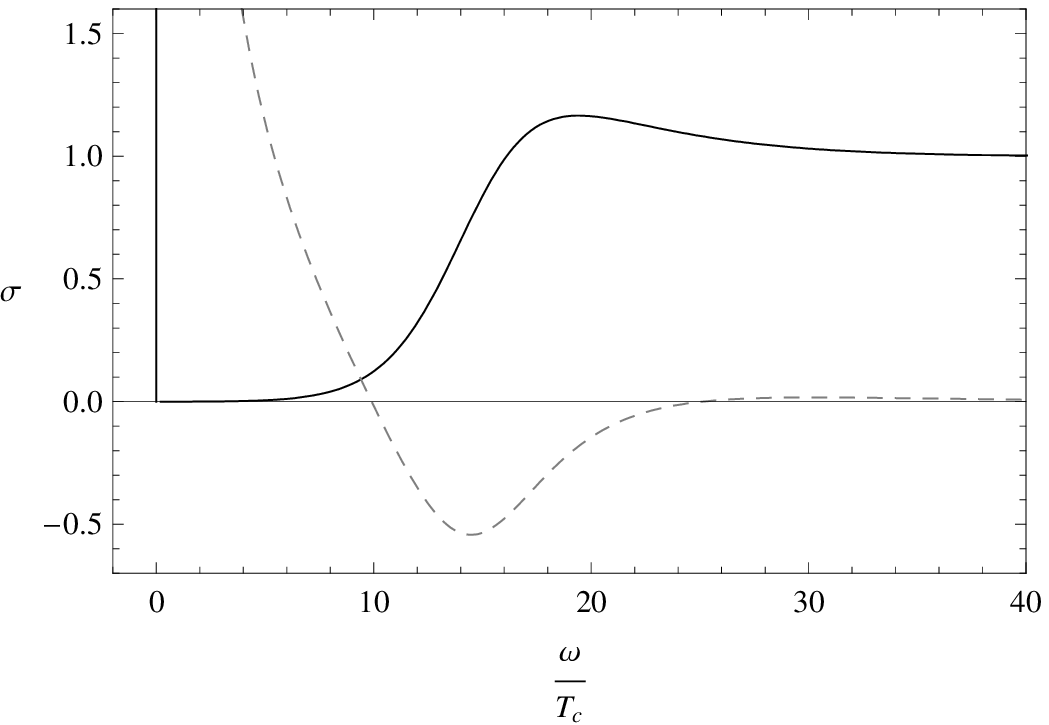}
\caption{ The conductivity for operator
$\langle\mathcal{O}_1\rangle$ at $T/T_c\approx0.160$ (up-left);
operator $\langle\mathcal{O}_2\rangle$ at $T/T_c\approx0.205$
(up-right) and operator $\langle\mathcal{O}_3\rangle$ at
$T/T_c\approx0.205$ (bottom). The solid curves denote the real
part of conductivity Re[$\si$], while the dashed curves the
imaginary part of conductivity Im[$\si$]. For three cases the real
parts all have a delta function at $\om=0$, which means the
infinite DC conductivity for superconductors.}
\end{figure}

\subsection{The conductivity for the cases of $m^2=-2/L^2$ and $m^2=0$}
\label{subsect:AA}

We calculate the conductivity for the cases of $m^2=-2/L^2$ and
$m^2=0$. The numerical results  are plotted in Fig.~3. For all
plots, the real part of the conductivity approaches to
Re[$\si$]$\rha 1$ when the frequency grows. This means the state
of the dual CFT becomes a normal one when $\om\rha\infty$. The
imaginary part of conductivity Im[$\si$] goes to infinity when the
frequency approaches zero. This phenomenon can be explained using
the Kramers-Kronig relation
 \be\label{kk}
 \text{Im}[\si(\om)]=-P\int^{\infty}_{-\infty}\frac{d\om'}{\pi}\frac{\text{Re}[\si(\om')]}{\om'-\om},\ee
where  $P$ denotes the principal value of the integration. Because
the real part has a delta function, Re$[\si(\om)]\sim \dl(\om)$,
 we can get from (\ref{kk}) that Im$[\si(\om)]\sim\frac{1}{\pi
\om}$, which means there exists a pole at $\om=0$.

We notice from the up-left plot of Fig.~3 that there is a spike in
the conductivity, which also appears for the  Schwarzschild-AdS
black hole case in~\cite{hr}. We will give a qualitative
interpretation for this spike later. Despite of the spike,
Re[$\si(\om)$] exponentially decreases when $\om\ll\om_g$.
Following the analysis in \cite{hr}, for
$\la>\la_{\text{BF}}=\frac 3 2$, we define $\om_g$  as the
frequency which minimizes the imaginary part of conductivity
Im[$\si$]. So for cases with $\la>\la_{\text{BF}}$ and
$T/T_c\approx 0.205$, we  find a roughly uniform ratio in Fig.~3,
that is,
 \be \frac{\om_g}{T_c}\approx 13,\ee
with the accuracy more than $93\%$. This  ratio is a little larger
than the one in \cite{hr}, but still in the same order.  In the
BCS theory,  the energy to break apart the condensate is about
$3.54T_c$ when $T=0$, our result $\om_g\approx 13T_c$ indicates
that in the \hl superconductor the energy is larger than the one
in the weakly coupled case.

It was also expected that when $\om\ll\om_g$, one has
 \be \text{Re}[\si]\sim e^{-\Delta/T},\ee
where $\Dl$ is a constant. This implies that when $T\sim 0$,
Re$[\si]$ should be strictly zero. However, as stressed in
\cite{hr2,kz}, Re$[\si(\om)]$ in fact depends on the frequency as
a power law form, {\it e.g.}, Re$[\si(\om)]\propto\om^{\dl}$ with
$\dl$ a function of some potential. Exploring the zero temperature
limit of this \hl superconductor would be an interesting issue.

  We can also see from Fig.~3 that when frequency is around $\om_g$,
Re$[\si(\om)]$ in the up-left plot grows slower than that in the
up-right one. This different growing type  may be expected from
type II and type I coherence factors~\cite{hhh}. Because type II
coherence will suppress absorption near $\om_g$, the real part of
conductivity will grow slower as the figure shows.

\subsection{Discussions on the Spike}
\label{subsect:BB}

Here we try to give a qualitative explanation for the spike
appearing in the
 up-left plot of Fig.~3. Following \cite{hr2}, we
 introduce a new radial variable $z$ as
  \be \label{newv} dz=\frac{dr}{f},\ee
It is easily seen that when $r\rha\infty$, $f\sim r^2$, therefore,
$z=-\frac1 r$ at the boundary. At the horizon, $f\rha 0$,
therefore, $z=-\infty$ is the black hole horizon in the $z$
coordinate. Using (\ref{newv}), we can rewrite  Eq.~(\ref{pert})
as
 \be \label{sch} -A_{x,zz}+V(z)A_x=\om^2 A_x.\ee
with \be V(z)=2q^2\psi^2f\ee
 Eq.~(\ref{sch}) is a one-dimensional Schr\"{o}dinger
 equation.  Then following \cite{hr2}, the conductivity can
 be expressed as
   \be \label{si}\si(\om)=\frac{1-\mathcal{R}}{1+\mathcal{R}}, \ee
 where $\mathcal{R}$ is the reflection coefficient. Naively we can
 deduce that if the barrier (potential $V(z)$) is very high and the
 incident energy of the incoming wave is lower than
 $V_{\text{max}}$, $\mathcal{R}$ will be close to $1$, which means
 $\si(\om)$ is very close to zero. From this analysis, $V_{\text{max}}\approx
 \om_g^2$. On the contrary, if $\om^2>V_{\text{max}}$, the incoming wave
 will easily cross over the barrier, therefore, $\mathcal{R}\rha0$
 and $\si(\om)\rha1$, which represents a normal state. The analysis
 above is consistent with what shows in Fig.~3.

 In the case of $\la=1$, the potential at the boundary is
 \be V(z)= 2q^2\psi^2 f\approx 2q^2(\psi^{(1)})^2,\ee
where $\psi^{(1)}$ is a finite constant. At the horizon, $f\rha 0$
and $\psi$ is finite, so $V(z)\rha0$ as $z\rha -\infty$. Because
$V(z)>0$ and it is continuous, we can deduce that $V(z)$ is
bounded from above, {\it i.e.}, $V(z)\leq V_{\text{max}}$.  The
potential can be plotted as Fig.4.
 \begin{figure}[htb]\label{poten}
\includegraphics[scale=0.76]{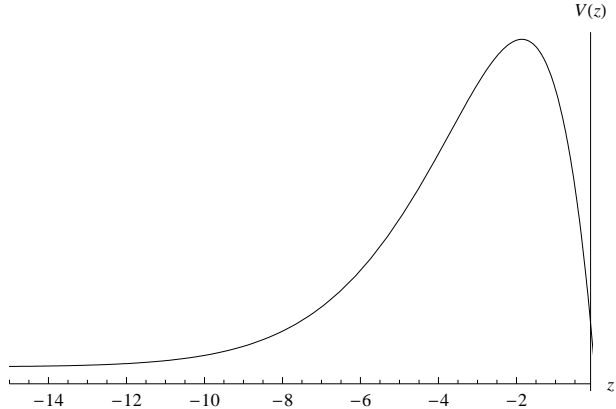}
\caption{The potential $V(z)$ which is bounded from above and goes
to zero at the horizon $z\rha-\infty$.}
\end{figure}
As is explained in \cite{hr2}, at low frequency the incoming wave
is almost entirely reflected. But if one low frequency wavelength
fits between the potential and the axis $z=0$, the reflected wave
can interfere destructively with the incoming wave and cause the
amplitude at $z=0$ to be exponentially small. According to the
relation (\ref{si}), this then produces a spike in $\si(\om)$. It
is suggested in~\cite{hr2} that by WKB approximation, there exists
$\om$ satisfying
 \be \int^0_{z_0}\sqrt{\om^2-V(z)}dz+\frac{\pi}{4}=n\pi \ee
for some integer $n$, and $V(z_0)=\om^2$, the spikes will appear.
 In our case, however, the potential $V(z)$ in the whole range of
$z$ is difficult to have an analytic expression  so that one is
not able to analytically fix the position of the spike.

\section{Concluding Remarks}

\label{sect:con}

 In this paper we have studied the phase transition of planar black holes in the \hl
 theory by introducing a Maxwell field and a complex scalar field in the probe approximation.
In particular,  we have discussed the cases of $m^2L^2=-2$
 and $m^2=0$, respectively. The former case shows that the two fallings of the complex scalar field
 are both normalizable modes. They are of scaling dimension one and two
 respectively. The later case has only a normalizable mode and the dual operator
 is of scaling dimension three. We found that the charged operator
 condensates behaving as $(T_c-T)^{1/2}$ , near the critical temperature,
 which is qualitatively consistent with the BCS
 theory. In particular, the condensates increase as the scaling dimension gets larger.
   We computed the
 electrical conductivities in the dual CFT when the temperature is
 much below the critical temperature. The dual CFT will turn to a normal state
 when the frequency is higher than the critical frequency $\om_g$. On the contrary, when
 the frequency is lower than $\om_g$, the real part of the conductivity will decrease
 exponentially.  There is an infinity DC conductivity at the zero
 frequency. In addition, it was found that when the scaling dimension of the charged
 operator is one, there exists a spike in the real part of the
 conductivity. We gave a qualitative explanation for this phenomenon using a
 one-dimensional Schr\"{o}dinger equation. We found that our
 results are quite similar to those in the case of Schwarzschild-AdS black
 holes. It implies that the holographic superconductivity observed
 in~\cite{gubser,hhh} is a robust phenomenon associated with
 asymptotic AdS black holes.

\begin{acknowledgments}\vskip -4mm
RGC would like to thank Kinki University, Osaka for warm hospitality
during his visit. HQZ would like to thank T. Chern, B. Hu, Z.Y. Nie,
S. Pu and Y. Zhou for helpful discussions. This work is supported
partially by grants from NSFC, China (No. 10535060, No. 10821504 and
No. 10975168) and a grant from MSTC, China (No. 2010CB833004).
\end{acknowledgments}

\end{document}